\documentclass[a4paper,11pt]{article}

\usepackage{graphicx}
\usepackage{amssymb}
\usepackage{amsmath}
\usepackage{amsfonts}
\newcommand{\imag}{{\rm Im}}

\def\dmatm{\Delta m^2_{@}}
\def\dmsol{\Delta m^2_{\odot}}
\def\bY{{\bf Y}}
\def\bR{{\bf R}}
\def\bU{{\bf U}}
\def\bM{{\bf M}}
\def\bmm{{\bf m}}

\newcommand{\diag}{{\rm diag}}

\begin{document}
\begin{flushright}
{DFPD-06/TH/11}
\end{flushright}
\vspace{0.5cm}

\begin{center}
{\huge \bf \sf \mbox{ A new bridge between leptonic CP violation}
and leptogenesis}\\[1cm]

{ {\large\bf G. C. Branco$^{a,\ast}$, R. Gonz\'alez Felipe$^{a,\S}$
and F. R. Joaquim$^{b,c,\ddag}$}}
\\[7mm]
{\it $^a$ Centro de F\'{\i}sica Te\'{o}rica de Part\'{\i}culas,
  Departamento de F\'{\i}sica,\\ Instituto Superior T\'{e}cnico,
  Av. Rovisco Pais, 1049-001 Lisboa, Portugal}\\[3mm]
{\it $^b$ Dipartimento di Fisica ``G.~Galilei'', Universit\`a di
Padova
I-35131 Padua, Italy}\\[3mm]
{\it $^c$ Istituto Nazionale di Fisica Nucleare (INFN), Sezione di
Padova, I-35131 Padua, Italy}
\\[1cm]
\vspace{-0.3cm}

{\tt  $^\ast$\,E-mail: \hspace*{-0.3cm}gbranco@mail.ist.utl.pt\\
$^\S$\,E-mail: \hspace*{-0.3cm}gonzalez@cftp.ist.utl.pt\\
$\!\!\!^\ddag$\,E-mail: \hspace*{-0.3cm}joaquim@pd.infn.it}

\vspace{1cm}

{\large\bf ABSTRACT}
\renewcommand{\baselinestretch}{1.5}
\end{center}

\begin{quote}
{\large\noindent Flavor effects due to lepton interactions in the
early Universe may have played an important role in the generation
of the cosmological baryon asymmetry through leptogenesis. If the
only source of high-energy $CP$ violation comes from the left-handed
leptonic sector, then it is possible to establish a bridge between
flavored leptogenesis and low-energy leptonic $CP$ violation. We
explore this connection taking into account our present knowledge
about low-energy neutrino parameters and the matter-antimatter
asymmetry observed in the Universe. In this framework, we find that
leptogenesis favors a hierarchical light neutrino mass spectrum,
while for quasi-degenerate and inverted hierarchical neutrino masses
there is a very narrow allowed window. The absolute neutrino mass
scale turns out to be $m \lesssim 0.1$~eV.}
\end{quote}

\renewcommand{\baselinestretch}{1.5}
\large

\section{Introduction}

The possibility of relating low-energy neutrino physics with the
baryon asymmetry of the Universe (BAU) produced via the mechanism of
thermal leptogenesis~\cite{Fukugita:1986hr} has received a great
deal of attention in the last few years~\cite{leptogenesis}. In the
simplest extension of the standard model (SM) where heavy neutrino
singlets are added to the particle content, light neutrino masses
arise through the seesaw mechanism~\cite{seesaw}. Besides providing
a natural mechanism to suppress neutrino masses, the seesaw
mechanism puts at our disposal the necessary ingredients to explain
the matter-antimatter asymmetry observed in our Universe. Indeed,
the out-of-equilibrium decays of heavy Majorana neutrinos, under the
presence of $CP$-violating interactions, produce a lepton asymmetry
which is partially converted into a baryon asymmetry by the
$(B+L)$-violating electroweak sphaleron
interactions~\cite{Kuzmin:1985mm}.

Recently, it has been noted that charged-lepton flavor effects play
a crucial role on the dynamics of the thermal leptogenesis
mechanism~\cite{Nardi:2005hs,Abada:2006fw,Abada:2006ea,Barbieri:1999ma}.
In particular, for temperatures below $\sim 10^{12}\,(10^9)$~GeV the
interactions mediated by the $\tau\,$($\mu$) are non-negligible and,
therefore, their effects should be properly taken into account in
the computation of the final value of the BAU. In the limit of
hierarchical heavy Majorana neutrinos $M_1 \ll M_{2} < M_3$, the
leptogenesis temperature is typically around $T\sim M_1$.
Consequently, depending on the actual value of $M_1$ and on which
charged-lepton Yukawa interactions are in equilibrium, one has
different possible scenarios.

In the one-flavor limit where all the charged leptons are equally
treated, one can show that a necessary condition for the mechanism
of leptogenesis to work is the presence of a nonvanishing
high-energy $CP$ violation in the right-handed neutrino sector. In
the flavored leptogenesis perspective, this remains true if $M_1
\gtrsim 10^{12}$~GeV, which corresponds to the temperature above
which all the charged lepton Yukawa interactions are out of
equilibrium. In this temperature regime, the $CP$ asymmetry
generated in the decays of the heavy Majorana neutrinos is summed up
over all flavors and its $CP$-violating part does not depend in
general on the low-energy $CP$-violating quantities which could be
potentially measured in future neutrino experiments. Therefore, in
the one-flavor approximation, the observation of low-energy leptonic
$CP$ violation does not necessarily imply the existence of a
nonvanishing BAU.

One may ask whether the above conclusions remain valid when flavor
effects are accounted for. In this letter we show that, in a general
class of models where $CP$ is an exact symmetry in the high-energy
right-handed neutrino sector, it is indeed possible to establish a
direct link between low-energy leptonic $CP$ violation and the
generation of the cosmological baryon asymmetry. In these models,
the baryon asymmetry only depends on the left-handed leptonic $CP$
phases, which in turn are determined by the low-energy Dirac and
Majorana neutrino phases~\cite{Abada:2006ea}. We shall also briefly
address the question on the possibility of naturally preserving $CP$
as a good symmetry of the right-handed neutrino sector.

\section{Leptogenesis and CP violation: a new perspective}

We work in the simple framework of the SM extended with three
right-handed neutrinos $N_i\, (i=1,2,3)$ with hierarchical heavy
Majorana masses $M_1 \ll M_{2} < M_3$. Working in the basis where
the charged-lepton Yukawa couplings and the heavy Majorana neutrino
mass matrix are diagonal, the relevant Dirac neutrino Yukawa
interaction is $\bY_{i\alpha} N_i \ell_\alpha H$, where
$\ell_\alpha\, (\alpha=e, \mu, \tau)$ are the SM lepton doublets and
$H$ is the Higgs doublet. We take advantage of the so-called
Casas-Ibarra parametrization~\cite{Casas:2001sr}
\begin{equation}
\label{CasasI}
\bY_{i\alpha}=\sqrt{M_i}\,\bR_{ik}\,\sqrt{m_k}\,\bU^\ast_{\alpha
k}/v\,,
\end{equation}
where $\bU$ is the low-energy leptonic mixing matrix, which
diagonalizes the effective neutrino mass matrix $\bmm_\nu$ in such a
way that
\begin{equation}
\label{diagon}%
\bmm_\nu=v^2\,\bY^T\,\bM^{-1}\,\bY=\bU^\ast\,\diag(m_1,m_2,m_3)\,\bU^\dag\;.
\end{equation}
Here $m_i$ are the effective light neutrino masses and $v \equiv
\langle H^0 \rangle \simeq 174$~GeV. The matrix $\bR$ in
Eq.~(\ref{CasasI}) is an orthogonal matrix, in general complex. In
what follows, we consider a class of seesaw models where $\bR$ is
real, corresponding to the cases where $CP$ is conserved in the
right-handed neutrino sector~\cite{Abada:2006ea}.

In this special case, the flavored $CP$ asymmetries generated in the
decays $N_1 \rightarrow \ell_{\alpha}\,H$ are simply given
by~\cite{Abada:2006ea}
\begin{align}
\label{flavCP}%
\varepsilon_\alpha = -\frac{3\,M_1}{16\,\pi
v^2}\frac{\sum_{k,j}\,m_k^{1/2} m_{j}^{3/2} \bR_{1k} \bR_{1j}\,{\rm
I}_{\alpha k j}}{\sum_k m_k \bR_{1k}^2}\,.
\end{align}
where ${\rm I}_{\alpha k j} \equiv \imag(\,\bU_{\alpha k}^\ast
\bU_{\alpha j} )$. Summing up over all flavors, $\varepsilon_1 =
\sum_\alpha \varepsilon_\alpha$, one recovers the standard
one-flavor result~\cite{Covi:1996wh}. It is straightforward to show
that if $\bR$ is real, then $\varepsilon_1=0$ due to the unitarity
of $\bU$. Thus, at temperatures where all lepton flavors are out of
equilibrium and the one-flavor approximation is valid, no lepton
asymmetry can be generated. This in turn implies an upper bound on
the lightest heavy Majorana neutrino mass, $M_1 \lesssim
10^{12}$~GeV, for the present scenario to be viable.

Clearly, the $CP$ asymmetries $\varepsilon_\alpha$ are very
sensitive to the type of light neutrino mass spectrum. Three
distinct cases are usually considered: hierarchical (HI), inverted
hierarchical (IH) and quasi-degenerate (QD) neutrinos,
\begin{align}
{\rm HI:}&\,\,m_1 \ll m_2\simeq (\dmsol)^{1/2}\;,\;m_3 \simeq (\dmatm)^{1/2}\,, \nonumber \\
{\rm IH:}&\,\,m_3 \ll m_1 \simeq m_2 \simeq (\dmatm)^{1/2}\,, \nonumber\\
{\rm QD:} &\,\,m \equiv m_1 \simeq m_2 \simeq m_3 >
(\dmatm)^{1/2}\,,
\end{align}
where $\dmsol = (7.9 \pm 0.6)\times 10^{-5}$~eV$^2$ and $\dmatm =
(2.6 \pm 0.4)\times 10^{-3}$~eV$^2$ are the \emph{solar} and
\emph{atmospheric} neutrino mass squared differences at $2\sigma$
level~\cite{Valle:2006vb}. The leptonic mixing matrix $\bU$ can be
parametrized in the form
$\bU_\delta\,\diag(1,e^{i\alpha},e^{i\beta})$, where $\bU_\delta$
contains the Dirac-type $CP$-violating phase $\delta$ and $\alpha,
\beta$ are Majorana phases~\footnote{For the matrix $\bU_\delta$ we
adopt the parametrization used in Ref.~\cite{Branco:2002ie}.}. For
the leptonic mixing angles $\theta_{ij}$ of the matrix $\bU_\delta$,
the presently available neutrino oscillation data yield $\sin^2
\theta_{12} = 0.30^{+0.06}_{-0.04}$, $\sin^2 \theta_{23} =
0.50^{+0.13}_{-0.12}$ and $s_{13}^2 \equiv \sin^2 \theta_{13} \leq
0.025$.

Using the orthogonality condition of the matrix $\bR$, one can show
that the $CP$ asymmetries (\ref{flavCP}) are bounded from above:
\begin{align}
\label{flavCPmax} {\rm HI:}\,\, |\varepsilon_\alpha|
&\leq\frac{3\,M_1\,(\dmatm)^{1/2}}{32\,\pi v^2}\,(1-\rho)\,|{\rm
I}_{\alpha 3 2}| \simeq 4 \times 10^{-7}\,|{\rm I}_{\alpha 3 2}|
\left(\frac{M_1}{10^{10}\,{\rm GeV}}\right)\,,
\nonumber\\
{\rm IH:}\,\, |\varepsilon_\alpha|
&\leq\frac{3\,M_1\,(\dmsol)^{1/2}}{32\,\pi v^2}\,\rho\,|{\rm
I}_{\alpha 2 1}| \simeq 1.5 \times 10^{-8}\,|{\rm I}_{\alpha 2 1}|
\left(\frac{M_1}{10^{10}\,{\rm GeV}}\right)\,,\nonumber\\
{\rm QD:}\,\, |\varepsilon_\alpha|
&\leq\frac{3\,M_1\,\dmatm}{64\,\pi v^2 m}\,(|{\rm I}_{\alpha 3
2}|^2+|{\rm I}_{\alpha 3 1}|^2)^{1/2} \nonumber \\ &\simeq 1.3
\times 10^{-7}\,(|{\rm I}_{\alpha 3 2}|^2+|{\rm I}_{\alpha 3
1}|^2)^{1/2} \left(\frac{M_1}{10^{10}\,{\rm GeV}}\right)
\left(\frac{0.1\,{\rm eV}}{m}\right)\,,
\end{align}
where $\rho=(\dmsol/\dmatm)^{1/2}\simeq 0.17$. It is interesting to
note that in the case of a real matrix $\bR$, the $CP$ asymmetries
$\varepsilon_\alpha$ vanish for exactly degenerate light neutrinos
(see Eq.~(\ref{flavCP})). Therefore, for QD neutrinos, the
quantities $\varepsilon_\alpha$ turn out to be suppressed by the
absolute neutrino mass scale $m$, contrarily to what occurs in the
case of a complex $\bR$, where the upper bound on the flavor
asymmetries is proportional to $m$~\cite{Abada:2006fw}. We also
remark that for a heavy Majorana mass $M_1 < 10^9$~GeV, the above
asymmetries will be typically too small to account for the BAU.
Thus, in the present framework, flavored leptogenesis could be
viable if
\begin{align}\label{M1range}
    10^9~\text{GeV} \lesssim M_1 \lesssim 10^{12}~\text{GeV}.
\end{align}
Since in this mass window only the $\tau$ Yukawa coupling is in
thermal equilibrium, the final value of the baryon asymmetry per
entropy density can be written as~\cite{Abada:2006ea}
\begin{align}\label{finalBAU}
    Y_B \equiv \frac{n_B}{s} = -\frac{12}{37} \left(\frac{115}{36}
    Y_2 +\frac{37}{9} Y_\tau\right)\,,
\end{align}
where $Y_2$ is a combined density coming from the indistinguishable
$e$ and $\mu$ asymmetries. The individual flavor densities $Y_2$ and
$Y_\tau$ can be found by solving the corresponding system of
Boltzmann equations. In the mass region (\ref{M1range}), it suffices
to consider the leptonic $CP$ asymmetry $\varepsilon_\tau$, since
$\varepsilon_2\equiv\varepsilon_e+\varepsilon_\mu=-\varepsilon_\tau$
when $\bR$ is real.

\begin{figure*}
\begin{tabular}{cc}
\includegraphics[width=7cm]{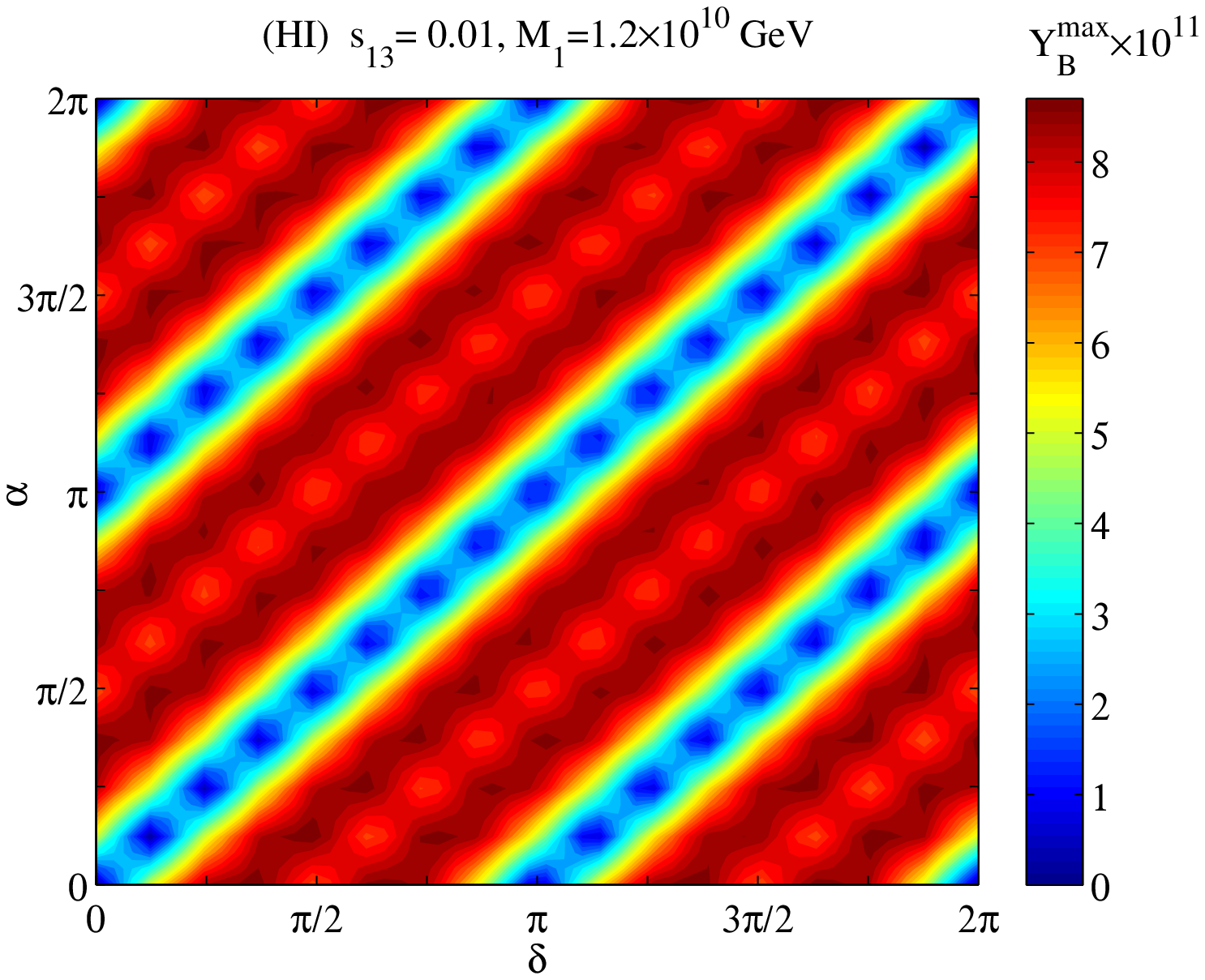}
&\includegraphics[width=7cm]{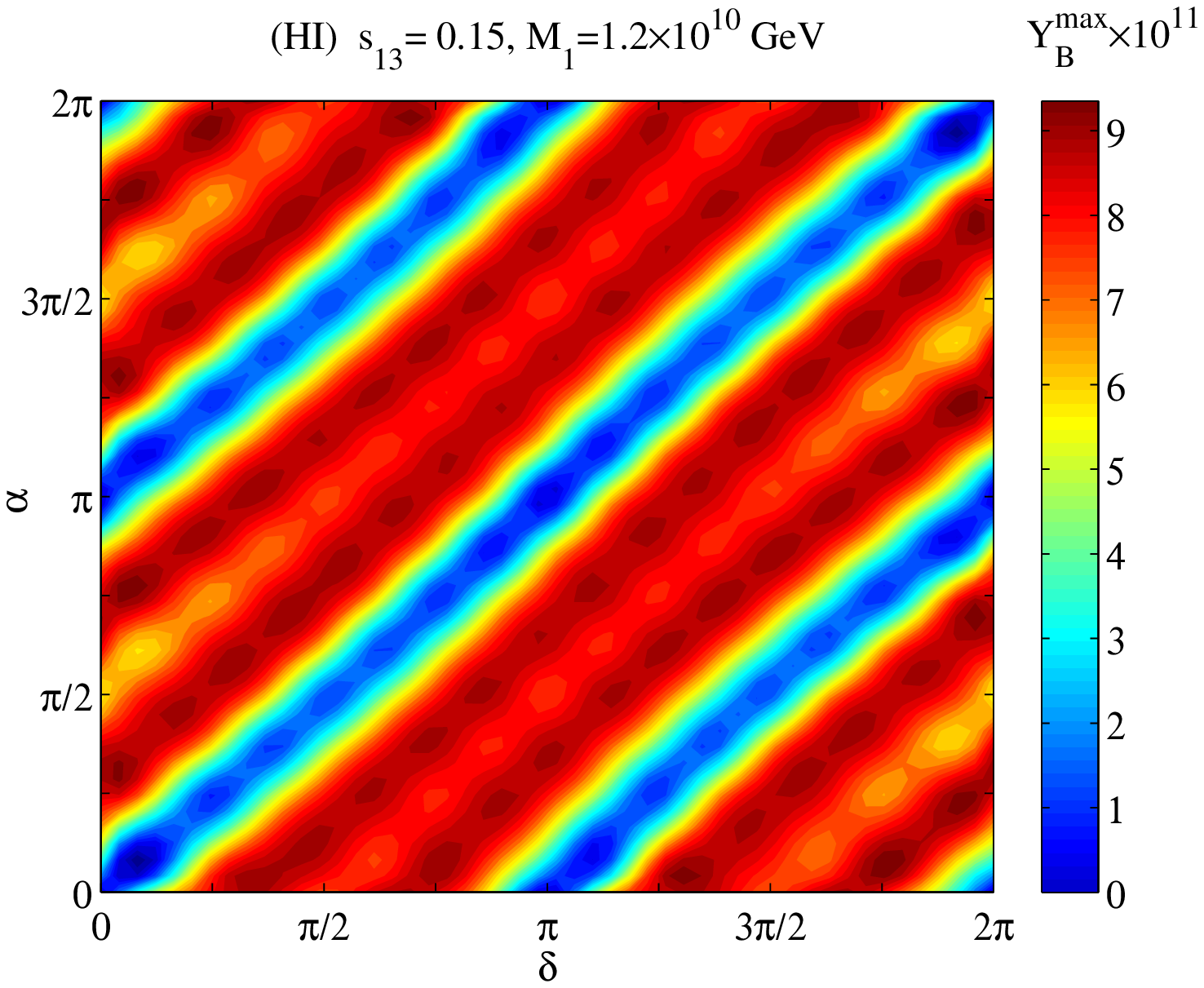}\\
\includegraphics[width=7cm]{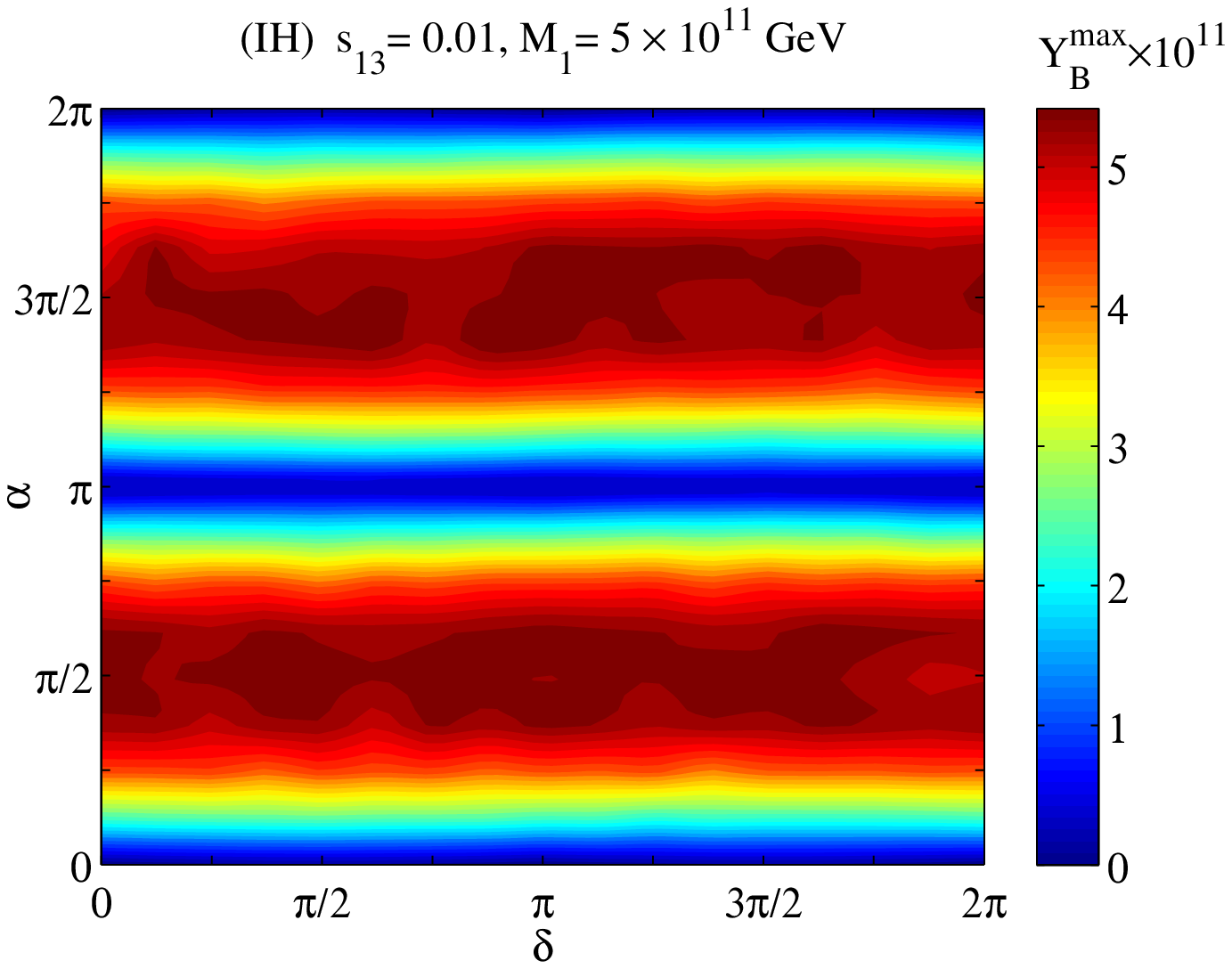} &
\includegraphics[width=7cm]{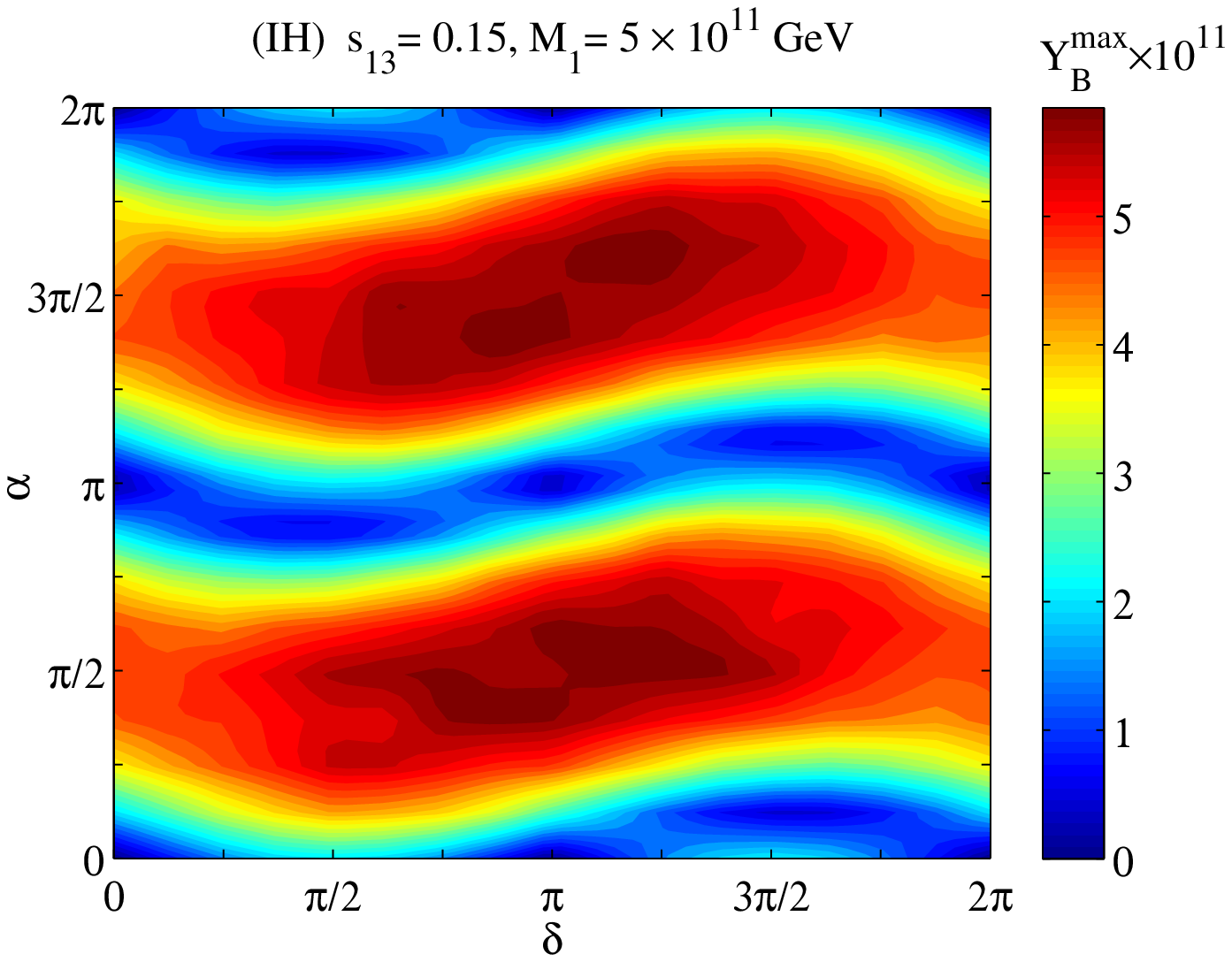}
\end{tabular}
\caption{(Color online) The correlation between the low-energy Dirac
($\delta$) and Majorana ($\alpha$) phases for different neutrino
spectra: normal hierarchy (HI) (upper panels) and inverted hierarchy
(IH) (lower panels). The color-shaded contours correspond to the
maximum of the baryon asymmetry that can be generated in models
where $CP$ is an exact symmetry of the right-handed neutrino sector.
We consider $s_{13}=0.01, 0.15$. The remaining low-energy neutrino
parameters are taken at their present central values.}\label{fig1}
\end{figure*}

\begin{figure}[t]
\begin{center}
\includegraphics[width=9cm]{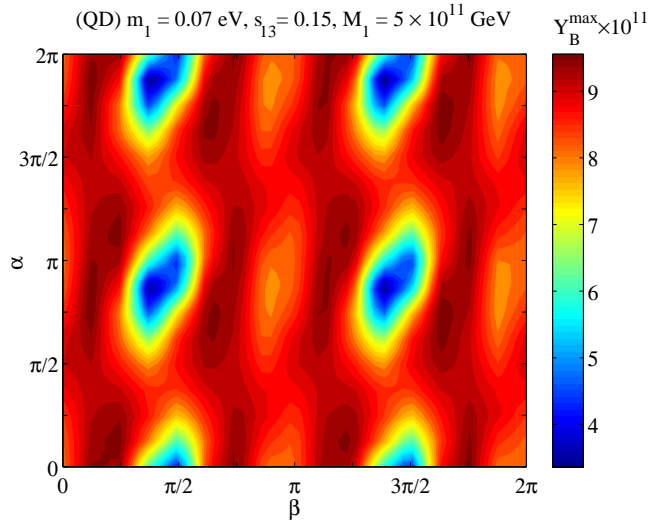}
\caption{(Color online) The correlation between the low-energy
Majorana phases, $\alpha$ and $\beta$, for the case of a
quasi-degenerate (QD) neutrino mass spectrum and maximal low-energy
Dirac $CP$ violation ($\delta=\pi/2$).}\label{fig2} \end{center}
\end{figure}

In Fig.~\ref{fig1} we present the correlation between the low-energy
Dirac ($\delta$) and Majorana ($\alpha$) phases for different
neutrino spectra: normal hierarchy (upper panels) and inverted
hierarchy (lower panels). The color-shaded contours correspond to
the maximum of the baryon asymmetry that can be generated in models
where $CP$ is an exact symmetry of the right-handed neutrino sector.
The curves were obtained by maximizing over all the possible values
of the orthogonal real matrix $\bR$. It is assumed in all cases that
right-handed neutrinos are not initially present in the thermal
plasma, but instead are created by inverse decays and scattering
processes. The plots are given for two different values of the
leptonic mixing angle $s_{13}$, namely, $s_{13}=0.01$ and $0.15$.
The remaining low-energy neutrino parameters are taken at their
present central values. A similar plot is presented in
Fig.~\ref{fig2} for the case of quasi-degenerate neutrinos. In the
latter case, the correlation between the low-energy Majorana phases,
$\alpha$ and $\beta$, is shown for maximal low-energy Dirac $CP$
violation, i.e. when $\delta=\pi/2$. We also remark that all the
results can be easily extrapolated to any value of $M_1$ in the
range of Eq.~(\ref{M1range}), since as can be readily seen from the
maximal $CP$ asymmetries in Eqs.~(\ref{flavCPmax}), there is
essentially a linear dependence between this parameter and the final
baryon asymmetry.

\begin{figure}
\begin{center}
\includegraphics[width=9cm]{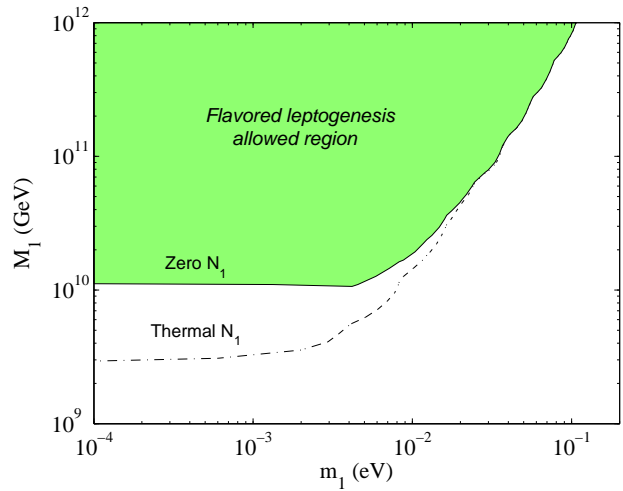}
\caption{(Color online) The region allowed for flavored leptogenesis
in the $(m_1,M_1)$-plane. The cases of zero and thermal initial
$N_1$ abundance are shown. The contour lines correspond to the lower
bound $Y_B = 8.4 \times 10^{-11}$.} \label{fig3}
\end{center}
\end{figure}

Comparing the results of Figs.~\ref{fig1} and \ref{fig2} with the
experimentally observed value $Y_B=(8.7\pm 0.3)\times
10^{-11}$~\cite{Spergel:2006hy}, it becomes clear that for IH and QD
neutrinos there is a very small allowed window for leptogenesis to
be viable in the present framework (i.e when no $CP$ violation
arises from the right-handed neutrino sector). Moreover, as expected
from the expression of the $CP$ asymmetries given in
Eq.~(\ref{flavCPmax}), we find an upper bound on the absolute
neutrino mass scale. In Fig.~\ref{fig3} we present the regions of
the $(m_1,M_1)$-plane where flavored leptogenesis can produce the
observed cosmological baryon asymmetry. The solid (dot-dashed) line
corresponds to a vanishing (thermal) initial $N_1$ abundance. From
the figure we obtain the lower bound $M_1 > 10^{10}$~GeV ($3\times
10^9$~GeV), as well as the upper bound $m_1 \lesssim 0.1$~eV.

\section{Conclusion}

We have studied the correlation between low-energy leptonic $CP$
violation and thermal leptogenesis in a class of seesaw models where
$CP$ is an exact symmetry of the high-energy right-handed neutrino
sector. In this case, and taking into account the role played by
flavor effects on the dynamics of the leptogenesis mechanism, one
can establish a correlation between low-energy $CP$ violation and
the cosmological baryon asymmetry. We have shown that for
hierarchical light neutrino masses a successful flavored
leptogenesis requires $M_1 > 10^{10}$~GeV ($3\times 10^9$~GeV) for a
zero (thermal) initial abundance of the lightest right-handed
neutrino $N_1$. On the other hand, values of $M_1$ close to the
upper bound of $10^{12}$~GeV (above which the present scenario is
not viable) are needed for the inverted hierarchy and
quasi-degenerate neutrino masses, thus leaving a very narrow window
for these neutrino mass spectra. We find an upper bound on the
absolute neutrino mass scale: $m \lesssim 0.1$~eV.

Regarding the correlation between the low-energy $CP$-violating
phases and the value of the BAU, our analysis has shown that there
are certain combinations of phases which are excluded for all values
of $M_1$ (see plots). Furthermore, in this class of models, the
observation of low-energy leptonic $CP$ violation would in general
indicate that the observed baryon asymmetry could have indeed been
created through the leptogenesis mechanism. Future information about
low-energy $CP$ violation either from neutrino oscillations or
neutrinoless double beta decay searches could further constrain the
present scenario.

Finally, we briefly address the question of how to construct, in the
seesaw framework, a model where $CP$ violation only occurs in the
left-handed neutrino sector. In general, once one allows for $CP$
violation through the introduction of complex Yukawa couplings, $CP$
arises both in the left-handed and right-handed sectors, so both
matrices $\bU_\delta$ and $\bR$ are complex. The simplest way of
restricting the number of $CP$-violating phases is through the
assumption that $CP$ is a good symmetry of the Lagrangian, only
broken by the vacuum. A model can actually be constructed, where one
has in a natural way $\bU_\delta$ complex while $\bR$ is real. Let
us consider the seesaw framework and impose $CP$ invariance at the
Lagrangian level. Now we introduce three Higgs doublets, together
with a $Z_3$ symmetry under which the left-handed fermion doublets
$\psi_{Lj}$ transform as $\psi_{Lj} \rightarrow e^{i 2\pi j/3}
\psi_{Lj}$ and the Higgs doublets as $\phi_j \rightarrow e^{-i 2\pi
j/3} \phi_j$, while all other fields transform trivially. It can be
readily shown that there is a region of parameters where the vacuum
violates $CP$ through complex vacuum expectation values $\langle
\phi_i^0 \rangle = v_i e^{i\theta_i}$. Due to the $Z_3$ restrictions
on Yukawa couplings, the combination $\bY \bY^\dagger$ is real, thus
implying a real $\bR$ (cf. Eq.~(\ref{CasasI})), but a complex
$\bU_\delta$ is generated. The drawback of such a scheme is that in
order to generate the required baryon asymmetry, leptogenesis would
have to take place at the TeV scale.

\vspace{0.5cm}

{\bf Acknowledgments:} We thank A. Riotto for private
communications. The work of R.G.F. and F.R.J. is supported by {\em
Funda\c{c}\~{a}o para a Ci\^{e}ncia e a Tecnologia} (FCT, Portugal) under the
grants \mbox{SFRH/BPD/1549/2000} and \mbox{SFRH/BPD/14473/2003},
respectively.  F.R.J. is also supported by INFN and PRIN Fisica
Astroparticellare (MIUR). F.R.J. thanks CFTP for hospitality during
the final stage of this work. The work of G.C.B. is supported by
CFTP-FCT UNIT 777 and POCTI/FNU/44409/2002, POCI/FP/63415/2005.

\vspace{0.5cm}

{\bf Note added:} While this work was in preparation, a related
preprint appeared~\cite{Pascoli:2006ie}, where some of the aspects
analyzed in this letter are also studied.

\end{document}